\documentclass[epj]{svjour}
\usepackage{psfig}
\begin{document}

\title{The $\eta{-}^3$He scattering length revisited}
\author{A. Sibirtsev\inst{^1}, J. Haidenbauer\inst{^1}, 
C. Hanhart\inst{^1} \and J.A. Niskanen\inst{^2}}

\institute{
Institut f\"ur Kernphysik, Forschungszentrum J\"ulich,
D-52425 J\"ulich, Germany \and
Department of Physical Sciences, PO Box 64, FIN-00014 University of
Helsinki, Finland }

\date{Received: date / Revised version: date}

\abstract{
The possible existence of $\eta$-mesic nuclei poses an interesting 
and still open issue of research. 
Since the occurence of such $\eta$-nucleus bound states is 
reflected in the corresponding $\eta$-nucleus scattering length, we 
critically review the present knowledge for the $\eta$$^3$He system. 
Specifically, we scrutinize the available experimental information
for the reaction $p{+}d \to\, \eta {+} ^3{\rm He}$ which is commonly used to 
extract the $\eta$$^3$He scattering length. We point out several striking 
discrepancies between the various measurements. Subject to those
inconsistencies we deduce a value of 
$a{=}|4.3{\pm}0.3|{+}i(0.5{\pm}0.5)$ fm. 
}
\maketitle

\section{Introduction}
The possible formation of $\eta$-nucleus quasibound states has been an 
interesting topic for a long time. However, so far no such states have 
been directly observed. It is also an open and heavily debated question 
what might be the lightest nuclei
for which such a bound state can occur.
For instance, investigations~\cite{Haider1,Liu,Chiang} based on 
optical models indicate carbon as the lower limit for 
nuclei able to bind an $\eta$ meson. Most recently~\cite{Haider2} this
limit was lowered to the $^4$He nucleus. In contrast, even formation of a
bound $\eta^3{\rm He}$ system is supported by other and different 
model calculations~\cite{Wycech,Belyaev1,Belyaev2,Rakityansky,Fix1}.
To clarify the situation experimental studies of this
system are proposed at GSI \cite{GSI1,GSI2} and COSY \cite{COSY1,COSY2,COSY3}. 
It is obvious that such experiments are very delicate and their design requires
good estimates for the relevant binding energies and widths of
$\eta$-mesic nuclei. Such estimates would dictate, for
instance, the necessary resolution of the detector and the 
required beam luminosity.

In this context very light nuclei are particularly interesting
because such systems are accessible to a microscopic treatment
whereas for heavier systems approximations have to be introduced
whose effects are difficult to quantify and, accordingly, might lead
to large uncertainties in the achieved results. 
In particular the $\eta^3{\rm He}$ system is very appealing
because it can be studied within the well-established 
Alt-Grassberger-Sandhas~\cite{Grassberger} and 
Faddeev -Yakubovsky~\cite{Faddeev} theories. 
The only but still very crucial ambiguity here is caused by our poor 
knowledge of the elementary $\eta$-nucleon ($\eta N$) interaction, which 
obviously enters any
microscopic calculation as an input. A compilation of values for the 
$\eta{N}$ scattering length, obtained from different $\eta{N}$ model
analyses, shows that its real part ranges from 0.20 to 1.05~fm, while the
imaginary part varies between 0.16 and 0.49~fm~\cite{Sibirtsev1}.
Since the elementary $\eta{N}$ interaction is not fixed, any $\eta$-nucleus
calculation \cite{Wycech,Belyaev1,Belyaev2,Rakityansky,Fix1} can only 
provide a range of results for the $\eta$-nucleus scattering lengths 
rather than a concrete prediction.

Under these circumstances it seems to be more promising to investigate 
a quantity closely related to the properties of the bound state, namely 
the $\eta$-nucleus scattering length \cite{Newton}. It is well-known
that in case of bound states the (real part of the) scattering length 
should be relatively large and negative. (We adopt here the sign
convention of Goldberger and Watson \cite{watson} common in meson physics.) 
Thus, studies of the $\eta$-nucleus interaction near
threshold can be used to determine the $\eta$-nucleus scattering
length, and then, in turn, would permit conclusions on the
existence of such $\eta$-nucleus bound states. Information
on the $\eta$-nucleus interaction can be deduced from analysing
the energy dependence of $\eta$ production reactions such as
$pd{\to}\eta^3{\rm He}$, $dd{\to}\eta^4{\rm He}$, etc. 
Certainly, the energy dependence of the production cross section
of those reactions itself is not sensitive to the sign of the real part
of the scattering length, but only to its magnitude, and therefore 
cannot provide direct evidence for the existence of a bound state.
(See, however, Ref. \cite{mix} for a possible experiment to determine
the sign of the real part.)
But even a good quantitative knowledge of the magnitude of the scattering 
length could already provide a strong hint for the existence of a 
$\eta$-mesic bound state and, more importantly, it would allow concrete
estimations for the energy range that should be scanned in dedicated
experiments. 

In the present paper we 
provide a systematic overview of the experimental
information available for the reaction $p{+}d{\to}^3{\rm He}{+}\eta$
\cite{Berger,Banaigs,Berthet,Mayer,Bilger,Betigeri,Kirchner}. 
In particular, we critically compare the results of the various measurements, 
which were partly performed for different kinematical conditions, in order 
to investigate the consistency of the data sets.
Special emphasis will be put on the data near the reaction
threshold which are commonly used to extract information about the 
$\eta ^3{\rm He}$ scattering length.  We discuss also results for the 
$\eta ^3{\rm He}$ scattering length that can be found in the literature.
The aim of the paper twofold. First we want to derive an new estimate for 
the $\eta ^3{\rm He}$ scattering length taking into account all available
low-energy data on the reaction $p{+}d{\to}^3{\rm He}{+}\eta$ 
and, secondly, we want to specify which further measurements 
are necessary in order to significantly improve the present situation. 

\section{Treatment of the final state interaction}

If a production reaction is governed by a strong $s$-wave interaction in 
the final state then, according to Watson and Migdal \cite{Watson,Migdal},
the energy dependence of the reaction amplitude
is basically determined by the 
on-shell scattering amplitude of the final state, i.e. by 
\begin{equation}
T(q) = \frac{1}{q \cot\delta -i q}\, .
\label{tfsi}
\end{equation}
where $q$ is the center-of-mass momentum of the strongly interacting particles 
in the final state and $\delta$ is the corresponding ($s$ wave) phase shift. 
Close to threshold the phase shift $\delta$ can be approximated 
by the effective range expansion
\begin{equation}
q \cot\delta = \frac{1}{a}+ \frac{r_0}{2}\, q^2\, ,
\label{efran}
\end{equation}
where for the $\eta^3{\rm He}$ interaction, of course, both the scattering 
length $a$ and the effective range $r_0$ are complex.

\begin{figure}[b]
\vspace*{-2mm}
\hspace*{-0.5mm}\psfig{file=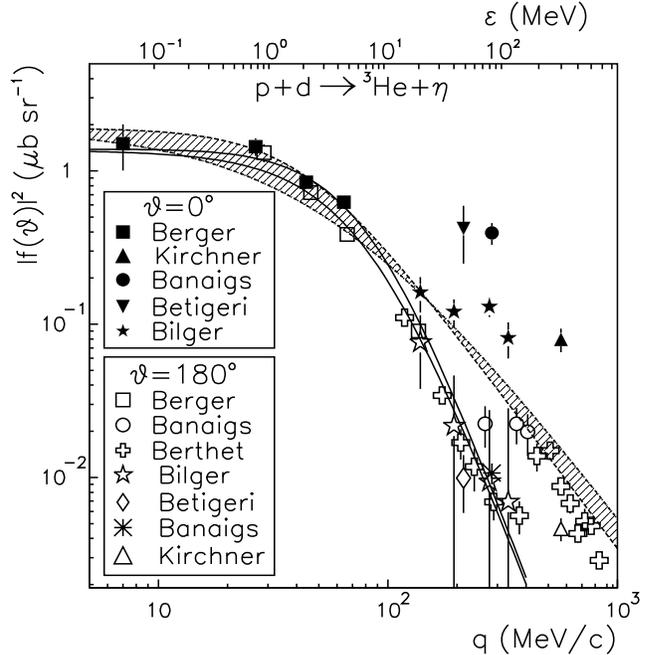,width=9.3cm,height=9.cm}
\vspace*{-6mm}
\caption{Spin averaged $p{+}d{\to}^3{\rm He}{+}\eta$ transition 
amplitude for forward (solid symbols) and backward (open symbols) 
$\eta$-meson production as a function of the final momentum $q$ in the 
c.m. system (lower axis) or excess energy $\epsilon$ (upper axis). 
The experiments are taken from Refs. \cite{Berger,Banaigs,Berthet}. 
In some cases \cite{Banaigs,Bilger,Betigeri,Kirchner} extrapolated 
results from a fit to the measured $\eta$-meson angular spectra are 
shown, cf. text. 
The solid lines are the fits of Ref.~\cite{Berger} to their 
$p{+}d{\to}^3{\rm He}{+}\eta$ data by Eq.~(\ref{fit1}) for
$\vartheta{=}0^o$ and $\vartheta{=}180^o$. 
The shaded area indicates results based on the correlation 
Eq.~(\ref{wilk}) reported in Ref.~\cite{Wilkin1}
for the extreme limits given by $\Re a{=}0$ and $\Im a{=}0$. 
Here $|f|^2$ was obtained by using Eq.~(\ref{fitn}).}
\label{etan6a}
\end{figure}

With the above sign convention a commonly quoted necessary
condition for the existence of a quasibound state is that $\Re a \ {<} \ 0$.
However, for having a quasibound state there is an additional
requirement, namely that the energy $E$ corresponding to the zero in
the denominator of Eq. (\ref{tfsi}) fulfils the relation  
$\Re E \ {<} \ 0$. This, in turn implies that 
\begin{equation}
\Re [a^3\, (a^\ast - r_0^\ast)] > 0
\end{equation}
in the two lowest orders in $r_0/a$. In the absence of the 
effective range term this reduces to the condition that 
$|\Re a|{>}\Im a{>}0$, given, e.g., in Ref.~\cite{Haider2}.

Neglecting terms of higher order
than $q^2$, the squared reaction amplitude, $|f|^2$, can be written as
\cite{Watson,Migdal}
\begin{eqnarray}
|f|^2&=& |f_p|^2 \cdot |T(q)|^2 
\nonumber \\
&\approx&\frac{|f_p|^2}{(1{+}\Im a q)^2{+}(\Re a q)^2{+}\Re a\Re r_0q^2
{-}\Im a\Im r_0q^2},
\label{fsqr}
\end{eqnarray}
where $f_p$ is the $s$-wave production operator, assumed to be independent
of the final momentum near the reaction threshold. 
 
While the coefficient of the term linear in $q$ is given by the
imaginary part of the scattering length alone, the $q^2$ term contains
{\it both the complex scattering length and effective range}. 
In case of the two-nucleon system $|a|{\gg}r_0$ and therefore 
a further approximation is reasonable. It consists in 
the neglect of the effective range, 
i.e. of the second term in Eq. (\ref{efran}), so that 
the reaction amplitude is simply given by 
\begin{equation}
f=\frac{f_p}{1-iaq}\, .
\label{fitm}
\end{equation}
Then Eq. (\ref{fsqr}) reduces to the form
\begin{equation}
|f|^2=\frac{|f_p|^2}{1{+}2 \Im a q {+}|a|^2 q^2}\, .
\label{fitn}
\end{equation}

However, for the $\eta^3{\rm He}$ system $a$ and $r_o$ are
expected to be of the same order of magnitude so that the above
approximation is not really justified. Thus, here $|a|^2$ as determined 
from Eq. (\ref{fitn}) can only be considered as an effective quantity
rather than the modulus of the physical scattering length. Clearly 
separating the scattering length and effective range is only possible by 
making further assumptions or within specific model calculations, which 
means in a model-dependent way.  
On the other hand, in the region very close to threshold where the
term linear in $q$ should dominate, in principle, there
is a possibility to determine the imaginary part of the
scattering length from the momentum dependence of the reaction  
amplitude $f(q)$. It should be feasible in the momentum range 
$q \le 1/2a$. 

\section{Data at low energies}

General information on the data base discussed in the present
paper is summarized in Table~\ref{exp}.

The application of the formalism described in the last section
is only sensible if two requirements are fulfilled: 
(i) the production data show a significant momentum
dependence near threshold; (ii) the production occurs 
predominantly in s-waves. 
A strong momentum dependence of the
spin averaged squared $p{+}d{\to}^3{\rm He}{+}\eta$ reaction amplitude
defined as
\begin{equation}
|f(\vartheta)|^2 := \frac{k}{q}\, \frac{d\sigma}{d\Omega},
\label{averm}
\end{equation}
was indeed seen in the first reported near-threshold measurement in 1988
\cite{Berger}. 
Here $k$ and $q$ are the initial and final particle momenta
in the center of mass system and $d\sigma{/}d\Omega$ stands for the
cms differential cross section. The measurements
were done only at the $\eta$-meson production angles $\vartheta{=}0^o$ and
$\vartheta{=}180^o$ in the cms and are presented in Fig.~\ref{etan6a} 
by full and open squares. The open circles (open crosses) in 
Fig.~\ref{etan6a} show earlier data of Banaigs {\it et al.}~\cite{Banaigs} 
(Berthet {\it et al.}~\cite{Berthet}) for $\eta$-meson production at 
$\vartheta{=}180^o$ at somewhat higher energies.

\begin{table}[t]
\caption{Data on the reaction $p{+}d{\to}^3{\rm He}{+}\eta$ discussed in
the present paper. $q$ is the cms momentum in the (final) $\eta ^3{\rm He}$
system and $\varepsilon$ is the corresponding excess energy. We use
$m_{^3{\rm He}}$ = 2809.414 MeV and $m_\eta$ = 547.3 MeV.}
\label{etadata}
\begin{tabular}{l|c|c|c|c}
\hline\noalign{\smallskip} 
&  Ref.  & Observable 
&  $q$ (MeV/c) & $\varepsilon$ (MeV)   \\
\noalign{\smallskip}\hline\noalign{\smallskip} 
 Berger & \cite{Berger} & $\!\!\!\!$
$ \ \sigma (0^o)$, $\sigma (180^o)$ $\!\!\!\!$
& 7{-}136 & 0.054{-}20 \\
Banaigs & \cite{Banaigs} & $\sigma (180^o)$ & 265{-}406 & 73{-}163 \\ 
Banaigs & \cite{Banaigs} & $\sigma (\vartheta)$ & 283 & 83 \\ 
Berthet & \cite{Berthet} & $\sigma (180^o)$ & 118{-}955 & 15{-}711 \\
Mayer & \cite{Mayer} & $\sigma_{tot}$, $A_{cm}$ & 11{-}75  & 0.13{-}6.11 \\ 
Bilger & \cite{Bilger}   & $\sigma_{tot}$, 
$\sigma (\vartheta)$ & 138{-}334 & 21{-}114 \\
Betigeri & \cite{Betigeri} & $\sigma_{tot}$, $\sigma (\vartheta)$ & 214 & 49 \\
Kirchner & \cite{Kirchner} & $\sigma_{tot}$, $\sigma (\vartheta)$ & 568 & 298 \\
\hline 
\end{tabular}
\label{exp}
\end{table}

Despite the discrepancies between the 
data~\cite{Banaigs,Berthet} on backward
$\eta$-meson production in the range $250{\le}q{\le}500$ MeV/c  
it is clear from Fig.~\ref{etan6a} that there is a strong $q$ dependence 
of $|f(\vartheta)|^2$ up to rather high energies.
Furthermore, from the data of Berger et al.~\cite{Berger}, 
which are available for both $\vartheta{=}0^o$ and $\vartheta{=}180^o$, 
one can conclude that the angular dependence is small for 
final momenta up to $q \approx 65$~MeV/c which suggests that 
the reaction amplitude should be dominated by the $s$-wave in this
momentum range. 

Further data on the reaction $p{+}d{\to}^3{\rm He}{+}\eta$ 
near threshold, obtained with the SPES2 spectrometer at Saclay,
were reported in 1996 \cite{Mayer}. This experiment provided
data on the total reaction cross section $\sigma_{\rm tot}$ and a 
forward-backward asymmetry in the cm system $A_{\rm cm}$ defined as
\begin{equation}
\frac{d\sigma}{d\Omega} = \frac{\sigma_{\rm tot}}{4\pi}[1 +
A_{\rm cm}\cos\vartheta] \ . 
\label{assym}
\end{equation}

\begin{figure}[b]
\vspace*{-6mm}
\hspace*{-1.mm}\psfig{file=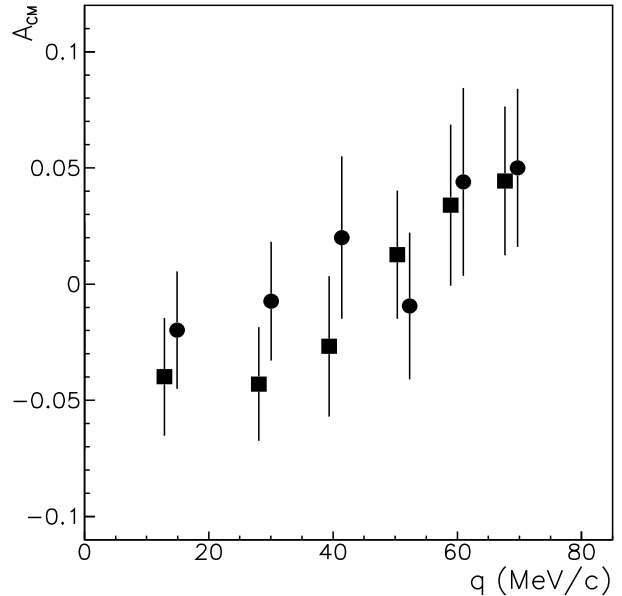,width=9.3cm,height=9.cm}
\vspace*{-7mm}
\caption{Data on the $\eta$-meson forward-backward
asymmetry $A_{\rm cm}$ as a function of final momentum $q$ 
from Ref. \cite{Mayer}. The different symbols show the experimental 
results obtained for different analyzing criteria \cite{Mayer}.}
\label{etan7a}
\end{figure}

The latter observable is
shown in Fig.~\ref{etan7a} as a function of the final momentum.
The full circles and squares indicate experimental results 
obtained under different criteria for data analysis.
Evidently, within 5\% accuracy in the amplitude the asymmetry is
consistent with zero. Thus, this measurement confirms that for 
$q{\le}70$~MeV the reaction $p{+}d{\to}^3{\rm He}{+}\eta$ is dominated by
the $s$-wave. Therefore, we will use the data on 
$p{+}d{\to}^3{\rm He}{+}\eta$ in this momentum range 
for investigating effects of the final state interaction 
(FSI) for the $\eta^3{\rm He}$ system. 

Assuming that the total reaction cross section is governed by the 
$s$ wave one can compute the average reaction amplitude squared, 
$|f|^2$, from the data of Ref. \cite{Mayer} by means of Eq.~(\ref{averm}). 
Corresponding results (now for the spin and angle averaged reaction
amplitude) are shown in Fig.~\ref{etan3b}. This figure contains also 
available data~\cite{Nikulin,Cameron,Roessle} for the reaction 
$p{+}d{\to}^3{\rm He}{+}\pi^0$. They are shown here in order to 
illustrate the strong momentum dependence of the $\eta^3{\rm He}$ 
channel, which is due to the corresponding FSI. 

\section{The $\eta^3{\rm He}$ scattering length}
Berger {\it et al.} \cite{Berger} did not attempt to extract the
$\eta^3{\rm He}$ scattering length from their data. However, they fitted
the data with the function
\begin{equation}
|f|^2=\frac{x}{(1 - yq\cos\vartheta + zq^2)^2} \ .
\label{fit1}
\end{equation}
The corresponding results are shown by solid lines in Fig.~\ref{etan6a} for
$\cos\vartheta{=}{\pm}1$. Note that Eq.~(\ref{fit1}) is
not the FSI correction to the production amplitude that 
follows from the Watson-Migdal approximation~\cite{Watson,Migdal}.
However, it can be matched to Eq. (\ref{fitn}) to order $q^2$ after 
averaging over the angle dependence. The explicit value for the modulus of 
the scattering length extracted in this way amounts to $|a|{=}3.4{\pm}0.1$~fm, 
using only the errors given in Ref.~\cite{Berger}.

\begin{figure}[t]
\vspace*{-3mm}
\hspace*{-1.mm}\psfig{file=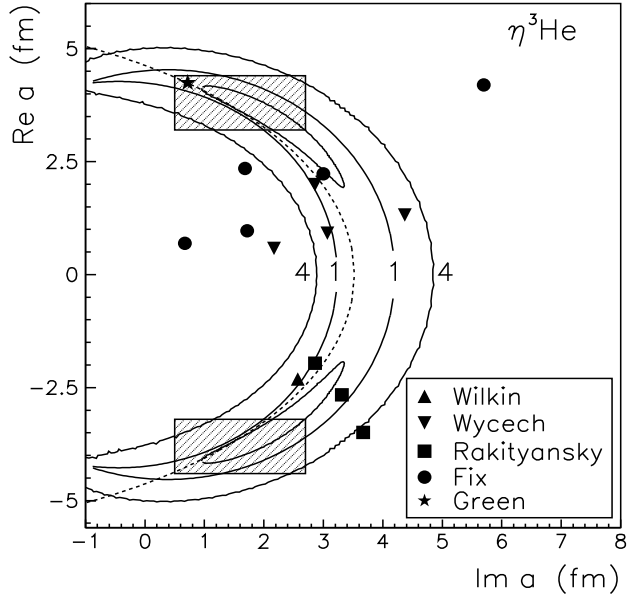,width=9.3cm,height=9.cm}
\vspace*{-7mm}
\caption{Real versus imaginary part of the $\eta^3{\rm He}$
scattering length. The shaded boxes indicate the value 
given by Mayer {\it et al.}~\cite{Mayer}. 
The solid contour lines show the result of our fit to the
data of Mayer {\it et al.}~\cite{Mayer} 
for $\chi^2{+}0.5$, $\chi^2{+}1$ and $\chi^2{+}4$ confidence levels,
respectively.
The dashed line is the parameterization of Eq.~(\ref{wilk}) from
Ref.~\cite{Wilkin1}. 
The symbols show results of various model calculations,
taken from Refs. \cite{Wycech} (inverse triangles), \cite{Rakityansky} 
(squares), \cite{Fix1} (circles), \cite{Wilkin1} (triangle) and
\cite{Green} (star). 
}
\label{etan9a}
\end{figure}

Employing Eq. (\ref{fitn}) Wilkin~\cite{Wilkin1} analysed
the preliminary Saclay data on the $p{+}d{\to}^3{\rm He}{+}\eta$ total
cross section \cite{Kessler}. He reported a correlation between the
real and imaginary parts of the $\eta^3{\rm He}$ scattering length 
in the form 
\begin{equation}
(\Re a)^2=21.44 - 0.449 (\Im a)^2 - 4.509 \Im a,
\label{wilk}
\end{equation}
as outcome of a $\chi^2$ minimization. 
This correlation is shown in Fig.~\ref{etan9a} by the dashed line. Note 
that Eq.~(\ref{wilk}) does not contain information about the
standard $\chi^2{+}1$ uncertainty of the fit and, because of the
unitarity condition, should be applied only for $\Im{a}{\ge}0$.

\begin{figure}[b]
\vspace*{-6mm}
\hspace*{-1.mm}\psfig{file=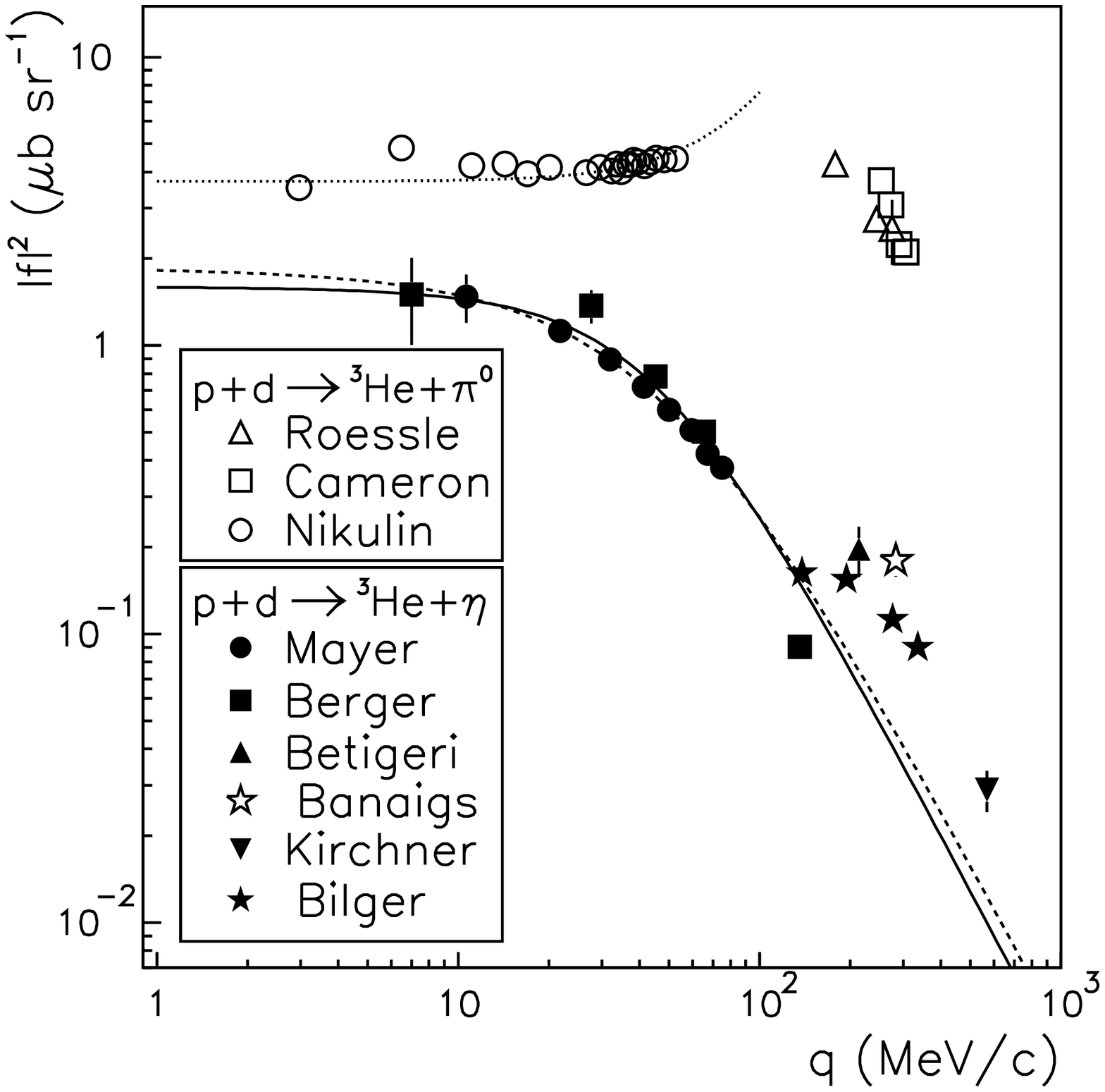,width=9.3cm,height=9.cm}
\vspace*{-5mm}
\caption{Spin and angle averaged transition amplitudes 
$|f|^2$ extracted from
$p{+}d{\to}^3{\rm He}{+}\pi^0$ \cite{Nikulin,Cameron,Roessle} and
$p{+}d{\to}^3{\rm He}{+}\eta$
\cite{Berger,Banaigs,Mayer,Bilger,Betigeri,Kirchner} data 
on total reaction cross sections as functions of the final momentum
$q$ in the c.m. system. The dotted line shows the fit to 
the reaction $p{+}d{\to}^3{\rm He}{+}\pi^0$ from Ref.~\cite{Nikulin}. 
The
solid line is our overall fit by Eq.~(\ref{fitm}) to low energy data published
by Mayer {\it et al.}~\cite{Mayer} and Berger {\it et al.}~\cite{Berger},
while the dashed line shows our fit to the data from
Mayer {\it et al.}~\cite{Mayer} alone.}
\label{etan3b}
\end{figure}

We took the $\eta^3{\rm He}$ scattering lengths constrained by 
Eq.~(\ref{wilk}) and employed Eq.~(\ref{fitn}) to calculate the average 
squared reaction amplitude as a function of the final momentum. 
Corresponding results are shown in Fig. \ref{etan6a}, where the
shaded area indicates the spread of $|f|^2$ with the limiting
scattering lengths of $a{\simeq}0{+}i3.51$ fm and $a{\simeq}4.63{+}i0$ fm 
as given in Eq.~(\ref{wilk}). It is worth mentioning 
that Wilkin did not include the data of Berger {\it et al.} in
his fit, though they were already available. 

Mayer {\it et al.} used also only their own data~\cite{Mayer} when 
they extracted the $\eta^3{\rm He}$ scattering length by utilizing 
Eq. (\ref{fitm}). Their result, 
\begin{equation}
a{=}|3.8{\pm}0.6|{+}i(1.6{\pm}1.1) \ \mbox{fm} \ ,
\label{sc1}
\end{equation}
is shown in Fig.~\ref{etan9a} by the shaded boxes. Since, as mentioned, 
the sign of the real part of the $\eta^3{\rm He}$ scattering length cannot
be inferred from a fit to the cross section data alone we include here 
boxes corresponding to $\pm {\Re} a$, with $a$ given by Eq. (\ref{sc1}).

The symbols in Fig.~\ref{etan9a} represent results of various
model calculations~\cite{Wycech,Rakityansky,Fix1,Wilkin1,Green} 
based on different approaches and different elementary $\eta N$ amplitudes. 
For convenience, selected results are also listed in 
Table~\ref{allowed} together with the elementary $\eta{N}$ scattering 
length that is employed in those model calculations.  Evidently, 
only the result from Ref.~\cite{Green} is in agreement with the 
$\eta^3{\rm He}$ scattering length extracted by Mayer 
{\it et al.}~\cite{Mayer}. 
In the course of our study we have refitted the data from Ref. \cite{Mayer}.
Our result is  
indicated by the solid contour lines in Fig.~\ref{etan9a} for $\chi^2{+}0.5$, 
$\chi^2{+}1$ and $\chi^2{+}4$ confidence levels. Apparently, 
it differs somewhat from the one published in Ref. \cite{Mayer}. 
Specifically, one can see that the error correlation matrix is not 
symmetric and that now several model predictions 
from the Refs.\cite{Wilkin1,Wycech,Rakityansky,Fix1,Green} lie within the 
$\chi^2{+}1$ confidence level.

We want to point out in this context that the value of the total $\chi^2$
at the minimum that results from our fit is $\chi^2{=}$0.16, which we find
to be much too low. Indeed for a statistically uncorrelated set 
of data points one would expect a value of 
$\chi^2$  = $N_{df} \pm \sqrt{2 N_{df}}$, where $N_{df}$ is the number
of degrees of freedom \cite{Nijmegen} -- which in this particular case 
would be 5. The error bars of the Saclay data are dominated by the 
statistical error \cite{Mayer} and, therefore, they cannot be the 
origin of this small $\chi^2$. Rather it seems to us that the published 
data points are simply not independent. 

\begin{table}[t]
\caption{Model calculations of the $\eta ^3{\rm He}$ scattering length 
that lie within the $\chi^2{+}1$ confidence level in Fig.~\ref{etan9a}. 
The employed values of the $\eta N$ scattering length and the used
approach is also specified.}
\label{allowed}

\begin{tabular}{l|c|c|l}
\hline\noalign{\smallskip}
Ref. & $ a(\eta ^3{\rm He})$ (fm) & $ a(\eta N) $ (fm)  & \,\,Comment \\
\noalign{\smallskip}\hline\noalign{\smallskip}
\cite{Wycech} & $1.99{+}i2.86$ & $0.48{+}i0.28$ & Multiple scattering \\
\cite{Wycech} & $0.92{+}i3.07$ & $0.43{+}i0.39$ & Multiple scattering \\
\cite{Rakityansky} & $-1.96{+}i2.86$ & $0.62{+}i0.30$ &
Finite-rank approx. \\
\cite{Rakityansky} & $-2.66{+}i3.31$ & $0.67{+}i0.30$ &
Finite-rank approx. \\
\cite{Fix1} & $2.23{+}i3.00$ & $0.57{+}i0.39$ & Faddeev-Yakubovsky \\
\cite{Wilkin1} & $-2.31{+}i2.57$ & $0.55{+}i0.30$ & Optical potential \\
\hline
\end{tabular}
\end{table}

\begin{figure}[b]
\vspace*{-6mm}
\hspace*{-1.mm}\psfig{file=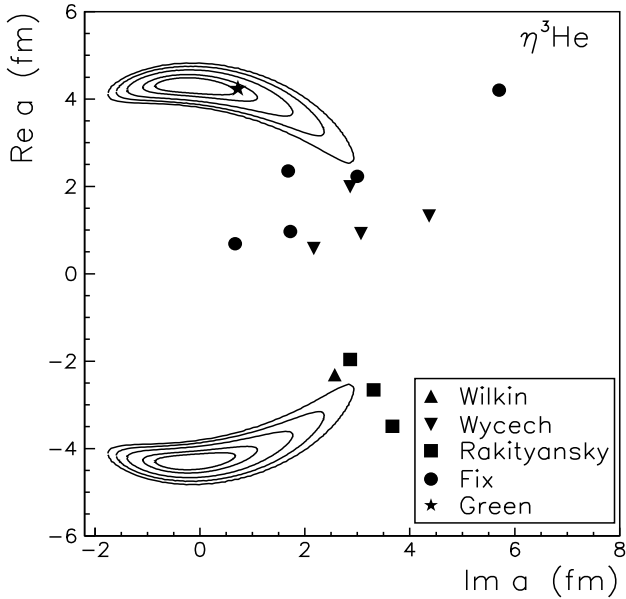,width=9.3cm,height=9.cm}
\vspace*{-7mm}
\caption{
Real versus imaginary part of the $\eta^3{\rm He}$
scattering length. 
The solid contour lines show the result of our fit to the
combined data of Mayer {\it et al.}~\cite{Mayer} 
and Berger {\it et al.}~\cite{Berger}
for $\chi^2{+}0.5$, $\chi^2{+}1$ and $\chi^2{+}4$ confidence levels,
respectively.
The symbols show results of various model calculations,
taken from Refs. \cite{Wycech} (inverse triangles), \cite{Rakityansky} 
(squares), \cite{Fix1} (circles),\cite{Wilkin1} (triangle) and 
\cite{Green} (star).}
\label{etan9}
\end{figure}

Considering this certainly to some extent strange feature of the
Saclay data one might {\it a priori} expect that an evaluation of the
$\eta^3{\rm He}$ scattering length from a combined data analysis
is very uncertain. Nevertheless we combine the data from Mayer
{\it et al.}~\cite{Mayer} and Berger {\it et al.}~\cite{Berger} to fit them
by Eq.~(\ref{fitm}). Corresponding result are presented in Figs.~\ref{etan3b} 
and \ref{etan9}. 
Besides yielding a substantially larger total $\chi^2{=}$57 
also the confidence contours are different for the combined fit
as can be seen by comparing Fig.~\ref{etan9} with Fig.~\ref{etan9a}.
(We show again the correlation between the real and imaginary part of
the $\eta^3{\rm He}$ scattering length for the $\chi^2{+}0.5$,
$\chi^2{+}1$, $\chi^2{+}2$, $\chi^2{+}3$ and $\chi^2{+}4$ confidence
levels.) Obviously, the combined analysis allows for a more definite
determination of the $\eta^3{\rm He}$ scattering length. 
In particular now the model predictions~\cite{Wycech,Rakityansky,Fix1,Wilkin1}
lie outside of the $\chi^2{+}1$ confidence level, except of the 
most recent result from Ref.\cite{Green}. On the other hand, 
the fact that the resulting $\chi^2$ minimum points to a 
$\eta^3{\rm He}$ scattering length with vanishing (or even slightly negative) 
imaginary part is definitely a reason to worry and is presumably a signal that 
the near-threshold data base is internally inconsistent and/or afflicted with
errors. Evidently, for further progress in the determination of the 
$\eta^3{\rm He}$ scattering length new measurements at final momenta 
$q{<}100$~MeV/c are required.

\section{Estimates for the imaginary part of the scattering length}
In principle, the imaginary part of the $\eta^3{\rm He}$
scattering length could be obtained from the total $\eta^3{+}{\rm He}$
interaction cross section $\sigma_{tot}$ by utilizing the optical theorem
in the limit $q \ \to \ 0$: 
\begin{equation}
\Im a = \lim_{q \to 0} \Im f_{tot}(\vartheta = 0^o) =  
\lim_{q \to 0} \frac{q}{4\pi}\,  \sigma_{tot} \ . 
\label{optic}
\end{equation}
Although $\sigma_{tot}$ is not accessible experimentally, one can use
at least experimental information on partial $\eta^3{+}{\rm He}$ reaction cross 
sections in order to deduce lower bounds on $\Im a$. This procedure works very
well for the $\eta N$ case where the magnitude of the imaginary part of the  
scattering length is strongly constrained by the data on the 
$\pi^- p \to  \eta n$ transition cross section \cite{Wilkin1,Sibirtsev1}. 

\begin{figure}[t]
\vspace*{-3mm}
\hspace*{0.mm}\psfig{file=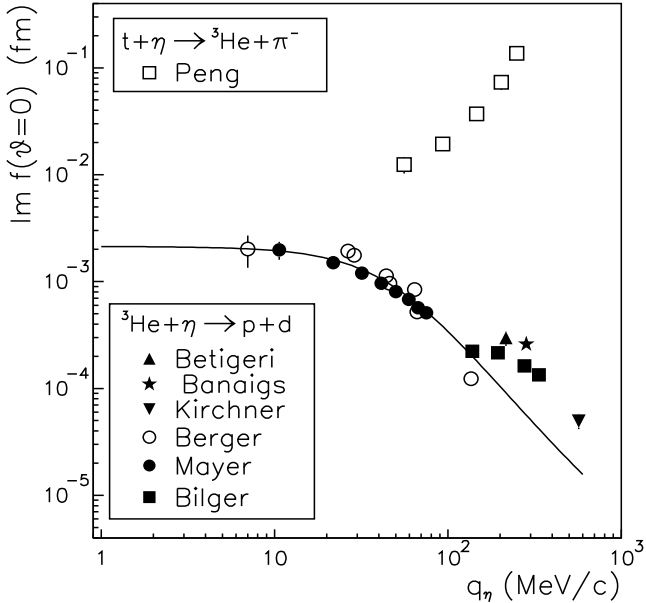,width=9.3cm,height=9.cm}
\vspace*{-5mm}
\caption{Bounds on the imaginary part of the $\eta^3{\rm He}$ forward 
scattering amplitude extracted from experimental results available for the
$p{+}d{\to}^3{\rm He}{+}\eta$
\cite{Berger,Banaigs,Mayer,Bilger,Betigeri,Kirchner} and 
$\pi^-{+}^3{\rm He}{\to}t{+}\eta$ \cite{Peng1,Peng2} reactions. 
The solid line shows our estimate based on an overall fit to low 
energy $p{+}d{\to}^3{\rm He}{+}\eta$ data.}
\label{etan3a}
\end{figure}

Using detailed balance
the $\eta^3{+}{\rm He}{\to}p{+}d$ cross section can be related to the data
available for the inverse reaction by 
\begin{equation}
\sigma (\eta{+}^3{\rm He}{\to}p{+}d) =\frac{3k^2}{q^2} \, 
\sigma (p{+}d{\to}^3{\rm He}{+}\eta) \ .
\label{bal1}
\end{equation}
In Fig.~\ref{etan3a} we show $\Im f(\vartheta = 0^o)$ 
obtained via Eqs.~(\ref{optic}) and (\ref{bal1}) from experimental
results~\cite{Mayer,Berger,Banaigs,Bilger,Betigeri,Kirchner} 
available for the reaction $p{+}d{\to}^3{\rm He}{+}\eta$. The solid 
line in Fig.~\ref{etan3a} shows the estimate based on our overall fit 
to the low energy data published by 
Mayer {\it et  al.}~\cite{Mayer} and Berger {\it et al.}~\cite{Berger}.

One can also evaluate the partial cross sections for 
$\eta{+}^3{\rm He}{\to}^3{\rm He}{+}\pi^0$ and
$\eta{+}^3{\rm He}\to {\rm t}{+}\pi^+$ 
from the data available for the reaction $\pi^-{+}^3{\rm He}{\to}t{+}\eta$ 
\cite{Peng1,Peng2}. Taking into account the
isotopical relations between the different reaction channels given by 
\begin{eqnarray}
\sigma (\eta{+}{\rm t}{\to}^3{\rm He}{+}\pi^-)= 
\sigma (\eta{+}^3{\rm He}{\to}{\rm t}{+}\pi^+)\nonumber \\
=2\sigma (\eta{+}^3{\rm He}{\to}^3{\rm He}{+}\pi^0)\, ,
\end{eqnarray}
the corresponding value of $\Im f(\vartheta = 0^o)$ can be estimated. 
It is also included in Fig.~\ref{etan3a}.

\begin{figure}[t]
\vspace*{-8mm}
\hspace*{-2.5mm}\psfig{file=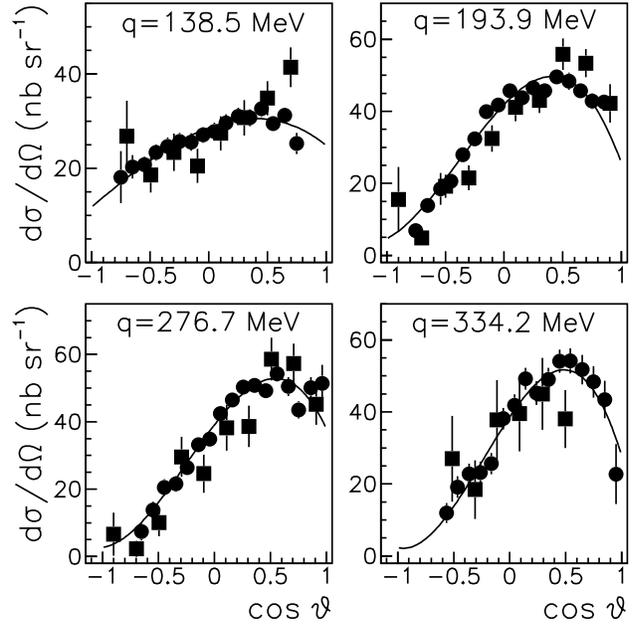,width=10cm,height=10.cm}
\vspace*{-7mm}
\caption{Angular spectra of $\eta$-mesons
produced in $p{+}d{\to}^3{\rm He}{+}\eta$ reaction at different
final momenta $q$. The data are from Ref. \cite{Bilger} where different 
symbols show results obtained with different analyzing criteria. 
The solid lines indicate the fit given in Ref.~\cite{Bilger}.}
\label{etan2}
\end{figure}

Unfortunately, 
the lower bounds for the imaginary part of the $\eta^3{\rm He}$
scattering length extracted from those reaction channels turn out to be rather
small, i.e. $\Im a{>}10^{-2}$~fm, and, therefore, are not very useful.
Presumably the bulk of the inelastic cross section comes from the reaction 
$\eta + ^3{\rm He} \to ppn$ which is, of course, not accessible experimentally.
In any case,
intuitively one would expect that $\Im a$ should be at least 3 times the
imaginary part of the elementary $\eta N$ scattering length. Indeed all 
results of microscopic calculations in the literature 
\cite{Wycech,Belyaev1,Belyaev2,Rakityansky,Fix1} (cf. also 
Table~\ref{allowed}) 
seem to be consistent with this hypothesis. Then, based on
the lower bound, $\Im a_{\eta N}{\approx}0.28$ fm, estimated by using the
optical theorem \cite{Wilkin1} one would arrive at $\Im a{\ge}0.84$ fm 
which might be a reasonable guess.

\begin{figure}[b]
\vspace*{-9mm}
\hspace*{-0.2mm}\psfig{file=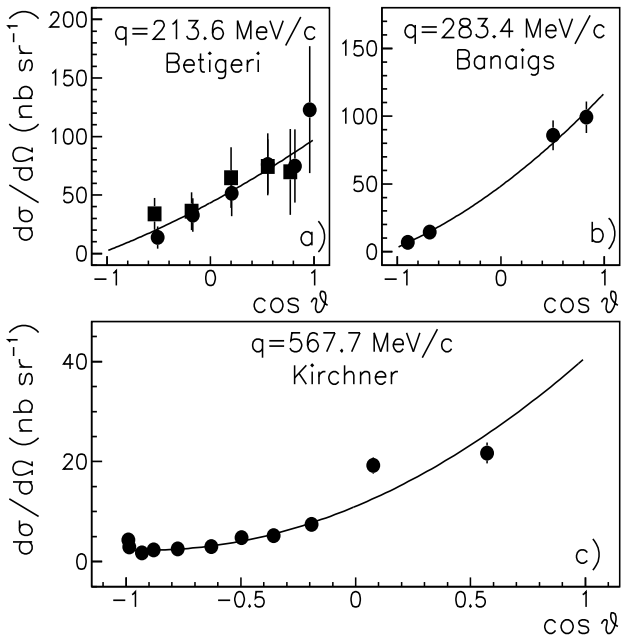,width=9.3cm,height=10.cm}
\vspace*{-8mm}
\caption{Angular spectra of $\eta$-mesons 
produced in the reaction $p{+}d{\to}^3{\rm He}{+}\eta$ at different
final momenta $q$. The data are from Refs. \cite{Banaigs,Betigeri,Kirchner}. 
The different symbols in (a) show results obtained with different analyzing
criteria~\cite{Betigeri}. The solid lines indicate our fit.}
\label{etan7}
\end{figure}

\section{Data at higher energies}

Angular spectra for the reaction $p{+}d{\to}^3{\rm He}{+}\eta$ 
at momenta $q{>}$100~MeV/c \cite{Bilger,Betigeri,Banaigs,Kirchner}  
are shown in Figs.~\ref{etan2} and \ref{etan7}. These data exhibit
already a strong asymmetry. Thus, it is clear that in this energy 
region the reaction is dominated by higher partial waves. 
Note that the solid lines in
Fig.~\ref{etan2} are taken from the original work while the curves in 
Figs.~\ref{etan7} show our own fit to the experimental results using 
Legendre polynomials.  

From those fits one can again compute the squared spin and
angle averaged transition amplitude and the corresponding results 
are included in Fig.~\ref{etan3b}. One can also extrapolate $|f|^2$ 
to very forward and backward angles and the corresponding values 
are shown in Fig.~\ref{etan6a}.
We detect substantial discrepancies between the 
$p{+}d{\to}^3{\rm He}{+}\eta$ forward cross sections 
extrapolated from the data of 
Bilger {\it et al.}~\cite{Bilger}, Banaigs {\it et al.}~\cite{Banaigs} and
Betigeri {\it et al.}~\cite{Betigeri}. The extrapolated data on backward
$\eta$-meson production~\cite{Bilger,Betigeri,Banaigs,Kirchner} are in
rough agreement with other published results~\cite{Berger,Berthet}
for $q{<}$300~MeV/c, taking into account that the data from Ref.~\cite{Bilger}
have large uncertainties.

\begin{figure}[t]
\vspace*{3mm}
\hspace*{-2.mm}\psfig{file=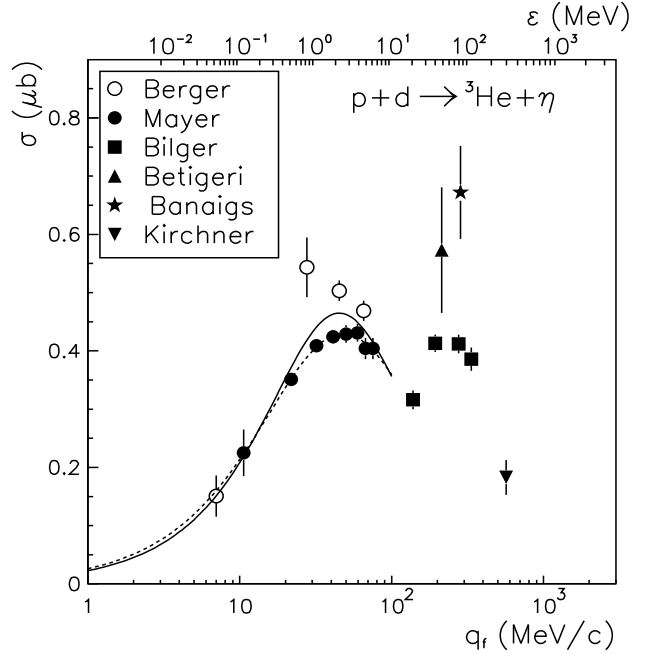,width=9.3cm,height=9.cm}
\vspace*{-7mm}
\caption{The $p{+}d{\to}^3{\rm He}{+}\eta$ total reaction cross
section as a function of the final momentum $q$ in the c.m. system (lower axis)
or excess energy $\epsilon$ (upper axis).  
The data are from Refs. ~\cite{Berger,Banaigs,Mayer,Bilger,Betigeri,Kirchner}.
The
solid line is our overall fit by Eq.~(\ref{fitm}) to low energy data published
by Mayer {\it et al.}~\cite{Mayer} and Berger {\it et al.}~\cite{Berger},
while the dashed line shows our fit to the data from
Mayer {\it et al.}~\cite{Mayer} alone.}
\label{etan1}
\end{figure}

The discrepancies between the available data are also reflected in 
Fig.~\ref{etan1}, where the experimental results 
\cite{Berger,Banaigs,Mayer,Bilger,Betigeri,Kirchner}
on the $p{+}d{\to}^3{\rm He}{+}\eta$ total reaction cross
section are shown as a function of the final momentum $q$ in the c.m. system
and the excess energy $\epsilon$. Here with open circles we also present
total reaction cross sections for the data of 
Berger {\it et al.}~\cite{Berger}, derived from their forward 
and backward $\eta$-meson production cross sections via Eq.~(\ref{assym}). 
Evidently, there is not much consistency between the various
data sets -- neither for low nor for higher energies.

\section{Summary}
We have critically reviewed the presently available data for the 
reaction $p{+}d{\to}^3{\rm He}{+}\eta$ with the aim of 
extracting the $\eta^3{\rm He}$ scattering
length. The experimental information on angular spectra clearly 
shows that the reaction is dominated by the $s$-wave up to final
momenta of around 70~MeV/c and, therefore, we have used all data in this 
energy range for the evaluation of the $\eta^3{\rm He}$ 
scattering length. 

The analysis provides strong indications that 
the low energy data published by 
Berger {\it et al}~\cite{Berger} and Mayer {\it et  al.}~\cite{Mayer} 
are not consistent with each other. 
The overall fit to all low energy data~\cite{Berger,Mayer} 
results in a large total $\chi^2$ of 57, however clearly locates the 
scattering length
as $\Im a{=}0.5{\pm}0.5$~fm and $\Re a{=}4.3{\pm}0.3$~fm. The fit to the
data from Mayer {\it et  al.}~\cite{Mayer} alone results in a rather 
small total $\chi^2$ of only 0.16, but yields an $\eta^3{\rm He}$ scattering
length with much too large statistical uncertainty.

Further progress
in the determination of the $\eta^3{\rm He}$ scattering length requires
new measurements at final momenta $q{\le}70$~MeV/c in order to settle the
ambiguities exhibited by the present data base. In particular, it would be
nice to obtain information about the $p{+}d{\to}^3{\rm He}{+}\eta$ 
reaction cross section very close to threshold, because this would provide 
us with more stringent constraints for the imaginary part of the 
scattering length. 
Furthermore, measurements of the angular spectrum of the $\eta$ meson
at energies corresponding to final momenta around $q = 70$~MeV/c 
would be very useful. Such data would allow to examine whether the reaction
is still dominated by the $s$-wave up to this energy -- as we assumed in
our analysis. The mentioned 
experiments could be done at accelerator facilities like COSY 
and CELSIUS \cite{COSY2,Khoukaz}. 

Our systematical analysis shows that the  
$p{+}d{\to}^3{\rm He}{+}\eta$ data available at larger final momenta, 
$q{>}$100~MeV/c, are also not consistent with each other. Obviously this
situation constitutes a substantial difficulty in the comparison between the 
experimental results and theoretical calculations. Here too the problem can be 
solved only by improving and expanding the data base for the reaction
$p{+}d{\to}^3{\rm He}{+}\eta$.

\begin{acknowledgement}
This work was performed in part under the contract No. DE-FG02-93ER40756
with the University of Helsinki. Financial support for this work was 
also provided in part by the
international exchange program between DAAD (Germany, Project No.
313-SF-PPP-8) and the Academy of Finland (Project Nos. 41926
and 54038).
\end{acknowledgement}

\end{document}